\documentclass[aps,prl,twocolumn,superscriptaddress,showpacs]{revtex4}
\usepackage[english]{babel}
\usepackage{amsmath}
\usepackage{graphicx}
\newcommand\bp[2]{(#1,#2)}%
\newcommand\cs{{c^*}}%
\newcommand\ie{\emph{i.\,e.\ }}%
\newcommand\fig{Fig.}%
\newcommand\kB{\mathrm{k_B}}%
\newcommand\Kt[1]{\kappa(#1)}%
\newcommand\Li[2]{\mathrm{Li}_{#1}\left(#2\right)}%
\newcommand\psp{\;}%
\newcommand\Tcr{{T_{\mathrm{cr}}}}%
\newcommand\vf[1]{{v_{\mathrm{f}}(#1)}}%
\newcommand\vl[1]{{v_{\mathrm{l}}(#1)}}%
\newcommand\wcr{{w_{\mathrm{cr}}}}%
\newcommand\zb{{z_{\mathrm b}}}%
\newcommand\zcr{{z_{\mathrm{cr}}}}%
\newcommand\Zg[1]{Z_{#1}}%
\newcommand\Zggt{\mathcal{Z}}%
\newcommand\Zgt{\tilde{Z}}%
\newcommand\Zh[2]{{Q_{#1}^{#2}}}%
\newcommand\Zht[2]{{\tilde{Q}_{#1}^{#2}}}%
\newcommand\zp{{z_{\mathrm p}}}%

\begin{document}
\title{Impact of loop statistics on the thermodynamics of {RNA} folding}

\author{Thomas R. Einert} \affiliation{Physik Department, Technische
  Universit\"at M\"unchen, 85748 Garching, Germany}%
\author{Paul N\"ager} \affiliation{Physik Department, Technische Universit\"at
  M\"unchen, 85748 Garching, Germany}%
\author{Henri Orland} \affiliation{Institut de Physique Th\'eorique, CEA Saclay,
  91191 Gif-sur-Yvette Cedex, France}%
\author{Roland R. Netz} \email[E-mail: ]{netz@ph.tum.de} \affiliation{Physik
  Department, Technische Universit\"at M\"unchen, 85748 Garching, Germany}

\date{\today}

\begin{abstract}
  Loops are abundant in native RNA structures and proliferate close to the
  unfolding transition.  By including a statistical weight $\sim \ell^{-c}$ for
  loops of length $\ell$ in the recursion relation for the partition function,
  we show that the heat capacity depends sensitively on the presence and value
  of the exponent $c$, even for a short explicit tRNA sequence.  For long
  homo-RNA we analytically calculate the critical temperature and critical
  exponents which exhibit a non-universal dependence on $c$.
\end{abstract}
\maketitle

Apart from its role as an information carrier, RNA has regulatory and catalytic
abilities\cite{book}.  Since this RNA functionality is mostly determined by its
three-dimensional conformation, the accurate prediction of RNA folding from the
base sequence is a central issue\cite{Tinoco1971}.  To a fairly good
approximation RNA folding can be separated into the formation of a secondary
structure, completely determined by the enumeration of all base pairs present in
a given sequence, and the tertiary structure formation which only operates on
the already existing secondary structural elements\cite{Thirumalai}.  This
constitutes a major simplification compared to the protein folding problem.
Since the folding of even short RNA molecules takes much longer than reachable
with all-atomistic simulations including explicit solvent, the more modest goal
of obtaining the most probable secondary structures based on experimentally
derived base-pairing and base-stacking free energies has been
pursued\cite{Zuker,McCaskill1990}.

Due to the high number of unpaired bases, loops are abundant in RNA even at low
temperatures.  Polymer theory predicts the configurational weight of a loop
consisting of $\ell$ bases to decay as $\ell^{-c}$ where the exponent $c$ is
universal and depends on the number of strands emerging from the
loop\cite{Duplantier1986}.  We formulate the RNA partition function including
the proper weight of loops using the same exponent~$c$ for terminal, internal,
as well as multi-loops\cite{Bundschuh2005,Gerland2001}.  For a homo-RNA in the
thermodynamic limit a folding transition is known to exist in the finite range
$2<c<2.479$\cite{Gennes1968,Mueller2002,Mueller2003}.  We analytically calculate
the $c$-dependent critical exponents of that transition.  Critical effects are
quite small which explains why they are not manifest in numerical
calculations\cite{Stella}.  On the other hand, the non-critical effects of
varying $c$ are pronounced, even for real finite-length RNA sequences.  We
numerically calculate the heat capacity of a yeast tRNA with 76 bases using
experimentally determined base pairing and stacking free
energies\cite{Mathews1999}.  At low temperatures the most probable structure
consists of a characteristic clover-leaf structure and thus includes a
multi-loop with four helices. Neglecting the loop statistics shifts the maximum
of the heat capacity by more than ${20\,\mathrm{K}}$, whereas including a
realistic exponent $c$ gives heat capacity curves that agree much better in
shape with experimental results\cite{Privalov1978}.

\begin{figure}
  \centering \includegraphics{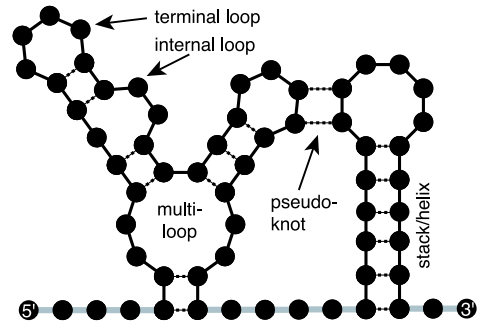}
  \caption{\label{fig:1} Schematic representation of a secondary RNA structure.
    Solid lines denote the RNA backbone, broken lines base pairs, and gray lines
    non-nested backbone bonds that are counted by the variable $M$; here
    $M=11$. }
\end{figure}

A primary RNA structure is fully determined by the base sequence $\{ b_N \} $
which is a list of nucleotides, $b_i=$C,G,A or U with $N$ entries.  In agreement
with previous treatments, a valid secondary structure is a list of all base
pairs with the constraint that a base can be part of at most one pair. In
addition, pseudo-knots are not allowed, \ie for any two base pairs $\bp{i}{j}$
and $\bp{k}{l}$ with $i<j$, $k<l$, and $i<k$ we have either $i<k<l<j$ or
$i<j<k<l$\cite{Zee}.  The statistical weight of a secondary structure depends on
the free energy of base pair formation but also on the stacking energy of
neighboring base pairs.  For two neighboring pairs $\bp{i}{j}$ and
$\bp{i+1}{j-1}$, the free energy containing both pairing and stacking is
$g[(b_i,b_j),(b_{i+1},b_{j-1})]$.  The statistical weight of a helical section
starting with $\bp{i}{j}$ and ending with $\bp{i+h}{j-h}$ is
$w^{\bp{i}{j}}_{\bp{i+h}{j-h}}=\exp\bigl[-\beta(g^{\mathrm{i}}[b_i,b_j]+
{\sum_{h'=1}^h g[(b_{i+h'-1},b_{j-h'+1}),(b_{i+h'},b_{j-h'})] +
g^{\mathrm{t}}[b_{i+h},b_{j-h}])} \bigr]$ where $\beta^{-1}=\kB T$. Here
$g^{\mathrm{i}}$, $g^{\mathrm{t}}$ are initialization and termination free
energies for base pairs located at the helix ends.  All values $g$,
$g^{\mathrm{i}}$, $g^{\mathrm{t}}$ are extracted from
experiments\cite{Mathews1999}, see supplementary information\cite{sup}.

In our notation, the canonical partition function $\Zh{i,j}{M}$ of a sub-strand
from base $i$ at the 5' end through $j$ at the 3' end depends on the number of
non-nested backbone bonds $M$\cite{Bundschuh2005,Mueller2002}, see
Fig.~\ref{fig:1}.  The recursion relations for $\Zh{i,j}{M}$ can be written as
\begin{subequations}\label{eq:1}
  \begin{equation}
    \label{eq:1a}
    \Zh{i,j+1}{M+1} = \frac{\vf{M+1}}{\vf{M}} \left[ \Zh{i,j}{M} +
      \sum_{k=i+M+1}^{j-N_{\mathrm{loop}}}\Zh{i,k-1}{M} \Zh{k, j+1}{0}\right]
  \end{equation}
  and
  \begin{multline}
    \label{eq:1b}
    \Zh{k,j+1}{0}= \sum_{h=1}^{(j-k-N_{\mathrm{loop}})/2}
    w^{\bp{k}{j+1}}_{\bp{k+h}{j+1-h}}\\ \times\sum_{m=1}^{j-k-1-2h}
    \Zh{k+1+h,j-h}{m}\frac{\vl{m+2}}{\vf{m}}\psp.
  \end{multline}
\end{subequations}
Eq.~\eqref{eq:1a} describes elongation of an RNA structure by either adding an
unpaired base (first term) or by adding an arbitrary sub-strand $ \Zh{k,
  j+1}{0}$ that is terminated by a helix.  Eq.~\eqref{eq:1b} constructs $ \Zh{k,
  j+1}{0}$ by closing structures with $m$ non-nested bonds, summed up in $
\Zh{k+1+h,j-h}{m}$, by a helix of length $h$.  $N_{\mathrm{loop}}=3$ is the
minimum number of bases in a terminal loop.  $\vf{M}$ and $\vl{M}$ denote the
numbers of configurations of a free and a looped chain with $M$ links,
respectively, for which we use the asymptotic forms $\vf{M} = {y}^MM^{\gamma-1}$
and $\vl{M} = {y}^MM^{-c}$~\cite{deGennes}.  The dependence on the monomer
fugacity $y$ and the exponent $\gamma$ drops out by introducing the rescaled
partition function $\Zht{i,j}{M}=\Zh{i,j}{M}/(y^{j-i} M^{\gamma -1})$ and will
not be considered further. The unrestricted partition function of the entire RNA
is given by $\Zg{N} = \sum_M \Zh{0,N}{M}$.

\begin{figure}
  \centering \includegraphics{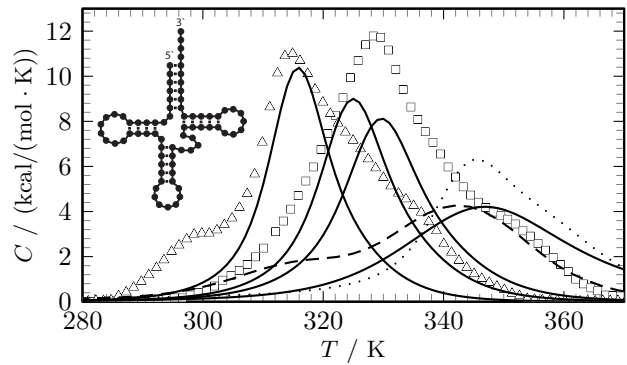}
  \caption{Experimental heat capacity of the tRNA-phe of yeast for NaCl
    concentrations ${20\,\mathrm{mM}}$ (triangles) and ${150\,\mathrm{mM}}$
    (squares)~\cite{Privalov1978}.  Solid lines show results using
    Eq.~\eqref{eq:1} with loop exponents $c=3.0,\ 2.16,\ 1.76,\ 0$ (from left to
    right), compared with the results from the program \texttt{RNAheat} in the
    \texttt{Vienna package}~\cite{Hofacker1994} which uses a linearized
    multi-loop entropy for large loops (dashed curve).  The dotted curve is
    obtained with $c=3$ and the same energy parameter set as for the solid
    curves, except for the loop initiation penalty which was omitted by setting
    $g^i=g^t$.  The inset sketches the low-temperature secondary RNA structure
    obtained from Eq.~\eqref{eq:1}, which perfectly matches experimental crystal
    structures.}
  \label{fig:2}
\end{figure}

The loop exponent is $c_{\mathrm{ideal}}=3/2$ for an ideal polymer and
$c_{\mathrm{SAW}}=d\nu \simeq 1.76$ for an isolated self avoiding loop with $\nu
\simeq 0.588$ in $d=3$ dimensions~\cite{deGennes}.  However, helices which
emerge from the loop increase $c$ even further.  In the asymptotic limit of long
helical sections renormalization group predicts $c_l = d\nu +
\sigma_l-l\sigma_3$ for a loop with $l$ emerging
helices~\cite{Duplantier1986,Kafri2000} where $\sigma_l=\varepsilon l(2-l) /16
+\varepsilon^2 l (l-2)(8l-21)/512 +\mathcal O (\epsilon^3)$ in an $\epsilon=4-d$
expansion.  One obtains $c_1=2.06$ for terminal, $c_2=2.14$ for internal loops
and $c_4=2.16$ for a loop with four emerging helices.  The variation of $c$ over
loop topologies that appear in the native structure of yeast tRNA-phe (shown in
the inset \fig~\ref{fig:2}) is thus quite small which justifies our usage of a
constant exponent $c$ for loops of all topologies.  For larger $l$ the
$\epsilon$ expansion prediction for $c_l$ becomes unreliable.  We therefore
treat $c$ as a heuristic input parameter which can be thought to account for
other loop-length dependent effects (such as salt-dependent electrostatic loop
self energies) as well.

We implement the recursion relation, Eq.~\eqref{eq:1}, numerically using a free
energy parameter set\cite{Mathews1999,sup} that allows for the wobble base
pair~GU in addition to the usual Watson-Crick pairs (GC and AU).  The boundary
conditions are $\Zh{i,j}{M}=0$ for $M>j-i$, $M<0$, or $j-i<0$, except for the
initial condition $\Zh{i,i-1}{-1}=1$.  In \fig~\ref{fig:2} we show the
experimental heat capacity of the tRNA-phe of yeast compared with our
predictions from Eq.~\eqref{eq:1} using $C=T \partial^2(\kB T \ln \Zg{N}) /
\partial T^2$.  The heat capacity peak corresponds to the gradual melting of the
secondary structure.  Although the RNA consists of just 76~nucleotides and is
therefore far from the thermodynamic limit where one expects asymptotic effects
to be important, the loop exponent~$c$ has drastic effects. Increasing $c$ from
0 to 3 changes the peak width and height and decreases the melting temperature
by more than ${30\,\mathrm{K}}$ (solid lines).  In similar studies on DNA, where
loops only appear close to the denaturation transition, the loop exponent was
found to have much less influence\cite{Blossey}.  In contrast, in RNA structures
a large fraction of bases are unpaired and the correct modeling of loops is more
important.  It is difficult to directly compare experimental and theoretical
curves, as the standard energy parameters used for secondary RNA-structure
predictions are determined at ${1\,\mathrm{M}}$ NaCl
concentration~\cite{Mathews1999}, while experimental heat capacity data is only
available at ${20\,\mathrm{mM}}$ and ${150\,\mathrm{mM}}$.  Although the actual
value of the loop exponent is not crucial (compare $c=1.76$ and $c=2.16$), the
effect of neglecting loop statistics (i.e. setting $c=0$) is almost as big as
omitting the loop initiation penalty contained in $g^i$ (dotted line), a well
established parameter, or changing the experimental salt concentration.  Current
secondary structure prediction tools approximate the entropy for large
multi-loops by an affine function $\ln (y^MM^{-c}) \approx \delta_0 +
\delta_1M$~\cite{Hofacker1994,McCaskill1990}, which corresponds to a loop
exponent $c=0$. This is corroborated by the near agreement of the results from
the Vienna package\cite{Hofacker1994} (broken line) with the results from
Eq.~\eqref{eq:1} using $c=0$.  It therefore is important to treat the
statistical weight of multi-loops on the same footing as terminal or internal
loops, if they appear in the RNA groundstate, as is the case for tRNA.  Our
formulation of the RNA partition function can be generalized to more complicated
loop weight functions to model the effects of salt or ligand binding.

We now consider homo-RNA, which can be realized experimentally with synthetic
alternating sequences like $[AU]_N$ or $[GC]_N$.  The goal is to extract the
critical asymptotic behavior embodied in Eq.~\eqref{eq:1} in the thermodynamic
limit. We neglect base stacking, helix initiation and termination and simply
give a statistical weight $w=\exp[-\varepsilon/(\kB T)]$ to each base pair.
This can be viewed as a coarse-graining approximation for natural or random RNA
above the glass transition.  Due to translational invariance Eqs.~\eqref{eq:1a}
and~\eqref{eq:1b} simplify and can be combined to
\begin{equation}
  \label{eq:2}
  \Zht{N+1}{M+1}=\Zht{N}{M} + w\sum_{n=M}^{N-1} \sum_{m=-1}^{N-n-2}\frac{
  \Zht{n}{M}\cdot\Zht{N-n-2}{m}}{(m+2)^{c}}\psp,
\end{equation}
where we introduced the total number of backbone segments $N=j-i$ of a part
which ranges from base number $i$ through $j$.  Next, we introduce the
generating function
\begin{equation}
  \label{eq:3}
  \Zggt(z,s) = \sum_{N=0}^\infty z^N\Zgt_N(s) =
  \sum_{N=0}^\infty\sum_{M=0}^\infty z^Ns^M\Zht{N}{M}\psp.
\end{equation}
For RNA with no external force one has $s=1$ and the sum over $M$, the number of
non-nested backbone bonds, is unrestricted. In general, $s=\exp(-\beta G(F))>1$
where $G(F)$ is the change in free energy of a non-nested bond caused by a
mechanical force $F$ applied at the RNA ends.  For a freely jointed chain one
has $G(F)=\beta^{-1} \ln\left[(\beta F a)^{-1} \sinh(\beta F a)\right]$, with
$a$ being the Kuhn length.  Combining Eqs.~\eqref{eq:2} and~\eqref{eq:3} yields
\begin{equation}
  \label{eq:4}
  \Zggt(z,s) = \frac{\Kt{z}}{1-sz\Kt{z}}\psp.
\end{equation}
where the function $\Kt{z} = \sum_{N=0}^\infty z^N \Zht{N}{0}$ is the grand
canonical partition function of an RNA with paired terminal bases.  From
Eqs.~\eqref{eq:3} and~\eqref{eq:4} $\Kt z$ follows as the positive root of
\begin{equation}
  \label{eq:5}
  \Kt{z} ( \Kt{z} -1) = w \Li{c}{z\Kt{z}}\psp,
\end{equation}
where we use the polylogarithm $\Li{c}{x} = \sum_{n=1}^\infty x^n/n^c$.  The
thermodynamic behavior for $N\rightarrow\infty$ is determined by the singularity
of the generating function $\Zggt(z,s)$ that is nearest to the origin in the
complex $z$-plane.  In particular, if $z^*$ is the dominant singularity with
$\Zggt(z,s) \sim C_1(z^* - z)^{\omega}$ for $z\rightarrow z^*$ and non-integer
$\omega$, then $\Zgt_N(s) \sim C_1^{-1} {N^{-(\omega +1)} z^*}^{-(N+1)}$ and the
Gibbs free energy becomes to leading order $\mathcal G = - \kB T \ln \Zgt_N = N
\kB T \ln z^*$.

\begin{figure*}
  \centering \includegraphics[width=\linewidth]{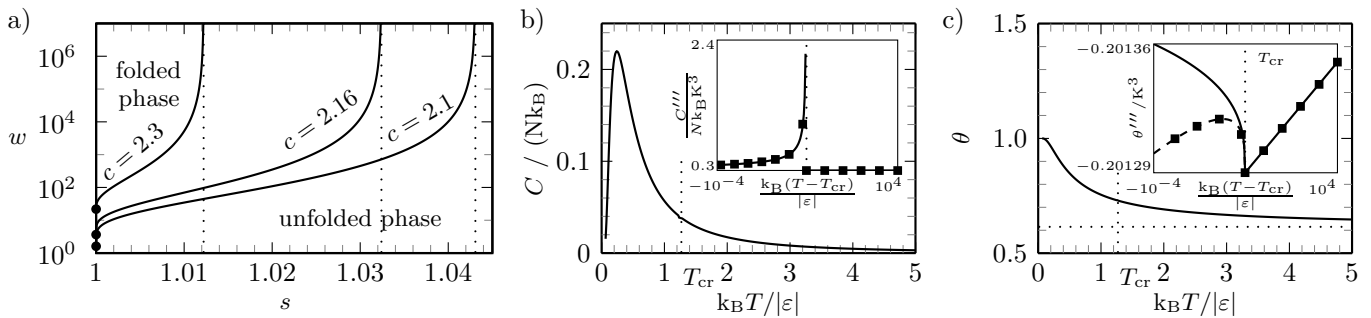}
  \caption{a) Phase diagram for three different values of the loop exponent~$c$
    as a function of the base pairing weight $w$ and force fugacity $s$
    featuring an unfolded phase (bottom) and a folded compact phase (top),
    following Eq.~\eqref{eq:8}.  The phase boundaries diverge at the vertical
    dotted lines.  For $c= \cs \simeq 2.479$ the phase boundary approaches $s=1$
    and therefore only the unfolded phase exists. For $c\leq2$ there is only the
    folded phase. The dots denote the unfolding transition in the absence of
    external force, \ie $s=1$, which is considered in b) and c): Temperature
    dependence of the b) specific heat $C$ and c) fraction of bound bases
    $\theta$ for $c=2.3$. The insets show the third derivatives $C'''=\mathrm
    d^3C/\mathrm d T^3$ and $\theta'''=\mathrm d^3\theta/\mathrm d T^3$ which
    clearly exhibit singular behavior.  Squares denote numerical solutions of
    Eqs.~\eqref{eq:5} and~\eqref{eq:6} ($T<\Tcr$) or~\eqref{eq:7} ($T>\Tcr$).
    The solid lines show the leading order expansion around $\Tcr$ (denoted by
    vertical dotted lines), according to which $C'''$ diverges with the exponent
    $\chi=2/3$ for $c=2.3$, see Eq.~\eqref{eq:10}, and $\theta'''$ is
    characterized by the exponent $\lambda=1/3$, see Eq.~\eqref{eq:13}.  In the
    inset of c) the analytical result including the next-leading order is also
    shown (broken line). The horizontal broken line in c) denotes the residual
    fraction of bound pairs at infinite temperature.}
  \label{fig:3}
\end{figure*}

It turns out that $\Zggt(z,s)$ has two singularities, first a branch point of
$\Kt{z}$ that follows by differentiating Eq.~\eqref{eq:5} and whose position
$\zb$ is determined by
\begin{equation}
  \label{eq:6}
  \Kt{\zb}^2 = w \Li{c-1}{\zb\Kt{\zb}} - w \Li{c}{\zb\Kt{\zb}}\psp.
\end{equation}
Second, a simple pole $\zp$ that follows from Eq.~\eqref{eq:4} and is determined
by
\begin{equation}
  \label{eq:7}
  s\zp\Kt{\zp} = 1\psp.
\end{equation}
The crossing of both singularities defines a critical point which is obtained by
solving Eqs.~\eqref{eq:5}-~\eqref{eq:7} simultaneously.  The critical base pairing weight
$\wcr$ as a function of the applied force fugacity $s$ reads in closed form
\begin{equation}
  \label{eq:8}
  \wcr = \frac{\Li{c-1}{s^{-1}} - \Li{c}{s^{-1}}} {\left(\Li{c-1}{s^{-1}} - 2
  \Li{c}{s^{-1}}\right)^2} \psp.
\end{equation}
In \mbox{\fig~\ref{fig:3}a} we show the phase diagram of RNA in terms of $w$ and
$s$ for different values of the loop exponent $c$.

Let us now consider the force-free case, \ie $s=1$. Eq.~\eqref{eq:8} simplifies
to $\wcr = \left({\zeta_{c-1} - \zeta_{c}}\right) {\left(\zeta_{c-1} - 2
\zeta_{c}\right)^{-2}}$ where $\zeta_{c} = \Li{c}{1}$ is the Riemann zeta
function.
It immediately follows that $\wcr$ is finite and non-zero only in the exponent
range $ 2<c<\cs $ with $\cs \simeq 2.479$ determined by $\zeta_{\cs -1} -
2\zeta_{\cs } = 0$.  For $c\rightarrow2$ from above, Eq.~\eqref{eq:8} predicts
$\wcr\rightarrow0$. Thus, for $c<2$, the RNA is always in the folded state and
$\Zgt_N(s) \sim N^{-3/2} \zb^{-N}$ is characterized by the branch point $\zb$
irrespective of how weak the pairing energy is~\cite{Gennes1968,Mueller2003}.
For $c> \cs $ the RNA is always unfolded and $\Zgt_N(s) \sim \zp^{-N}$ and is
determined by the simple pole.  Right at the critical point, \ie for $ 2<c<\cs
$, $\zcr=\zb=\zp$ and $w=\wcr$, the loop statistics become crucial and we obtain
the new scaling
\begin{equation}
  \label{eq:9}
  \Zgt_N(s) \sim N^{(2-c)/({c-1})} \zcr^{-(N+1)} \psp.
\end{equation}
This gives rise to non-universal critical behavior.  The specific heat possesses
a weak non-analyticity at the (n+2)-order critical point, meaning that the
$n^{\mathrm{th}}$ derivative with respect to temperature diverges as
\begin{equation}
  \label{eq:10}
  \frac{\mathrm{d}^n}{\mathrm{d} T^n}C \sim |\Tcr - T|^{-\chi} \psp,
\end{equation}
with $\chi =n- (3-c)/(c-2)$ for $T<\Tcr$ and $\chi = 1$ for $T>\Tcr$ and $n$
being the integer with $(c-2)^{-1}-1<n<(c-2)^{-1}$, see \mbox{\fig~\ref{fig:3}b}
and the supplementary information\cite{sup}. For $c\rightarrow \cs $ we have $n
\rightarrow 2$; for $c\rightarrow2$ we find $n\rightarrow\infty$. The fraction
of paired bases is obtained \emph{via} differentiation $ \theta =
\partial\ln\Zggt /(N\partial\ln w)$ and reads
\begin{equation}
  \label{eq:11}
  \theta = \frac{2 \Li{c}{\zb \Kt{\zb}}}{\Li{c-1}{\zb\Kt{\zb}}}\psp, \quad
  \text{for $T<\Tcr$ and}
\end{equation}
\begin{equation}
  \label{eq:12}
  \theta = 1-\left(1+4w\zeta_c\right)^{-1/2}\psp, \quad \text{for $T>\Tcr$.}
\end{equation}
As before the singularity at the critical point is very weak and the
$n^{\mathrm{th}}$ derivative of $\theta$ exhibits a cusp
\begin{equation}
  \label{eq:13}
  \frac{\mathrm{d}^n}{\mathrm{d} T^n}\theta \sim |T-\Tcr|^\lambda\psp,
\end{equation}
with $\lambda = (c-2)^{-1}-n$ for $T<\Tcr$ and $\lambda=1$ for $T>\Tcr$, see
\mbox{\fig~\ref{fig:3}c} where these asymptotic results are compared with
numerical solutions of Eqs.~\eqref{eq:5}-\eqref{eq:7}.

We account for the asymptotic statistics of loops by including a loop-length
dependent weight $\ell^{-c}$ in the recursion relation for the partition
function of RNA secondary structures. As a function of the loop exponent $c$ we
obtain exact critical exponents and boundaries between folded and unfolded
phases for a simplified homo-RNA model. Because the folding transition is at
least of fourth order, the singular contribution to thermodynamic observables
such as heat capacity or fraction of paired bases turns out to be quite
small. On the other hand, the non-singular contribution at temperatures well
below criticality depends crucially on $c$.  This is demonstrated for an
explicit sequence of a yeast tRNA by calculating heat capacities for various
values of $c$ and comparing with experimental data. It is seen that including
realistic values for $c$ is important and produces effects that are comparable
to changing base pair stacking parameters or changing the salt concentration.
So the conclusion is that while the dependence of critical properties on the
loop exponent $c$ is experimentally difficult to access and therefore largely
irrelevant, the dependence of non-critical properties on $c$ is important.

We are currently expanding the theory to allow for loop exponents that depend on
the actual number of helices emerging from a given loop and to include tertiary
contacts such as pseudo-knots or base triples which have been shown to play an
important role in RNA folding\cite{Stella}.

Support from the \emph{Elitenetzwerk Bayern} within the framework of
\emph{CompInt} is acknowledged.

\end{document}

\title{Supplementary material for: ``Impact of loop statistics on the
  thermodynamics of {RNA} folding''}

\author{Thomas R. Einert}
\affiliation{Physik Department, Technische Universit\"at M\"unchen - 85748
  Garching, Germany}%
\author{Paul N\"ager} \affiliation{Physik Department, Technische Universit\"at
  M\"unchen - 85748 Garching, Germany}%
\author{Henri Orland} \affiliation{Institut de Physique Th\'eorique, CEA Saclay,
  91191 Gif-sur-Yvette Cedex, France}%
\author{Roland R. Netz} \email[E-mail: ]{netz@ph.tum.de} \affiliation{Physik
  Department, Technische Universit\"at M\"unchen - 85748 Garching, Germany}

\date{\today}

\begin{abstract}
  We give the coefficients of the expansion of the branch point position $\zb$
  and the value $\Ktb$ of the function $\Kt{z}$ in the vicinity of the critical
  point.
We also list the free energy parameters used in our numerical study of the melting curve of the
yeast   tRNA-phe.
\end{abstract}

\maketitle

\section{Expansion of the branch point near the critical point}

For $T<\Tcr$ the dominant singularity of $\Zggt{}{}(z)$, which determines the thermodynamics, is
the branch point at $z=\zb$. The branch point has its origin in
the singular behavior of the function of $\Kt{z}$, which is determined by the
equation
\begin{equation}
  \label{eq:1}
  \Kt{z}(\Kt{z} - 1) = w \Li{c}{z \Kt{z}}\psp.
\end{equation}
In this section we are considering the asymptotic behavior of the branch point position
$\zb(T)$ and the functional  value $\Ktb(T):=\Kt{\zb(T)}$ at the branch point  in
the vicinity of the phase transition, in particular for $T\rightarrow\Tcr$ while 
$T<\Tcr$. The existence of the phase transition implies for the loop exponent
$2<c<c^*\approx2.479$.

The equation determining the position $\zb$ of the branch point follows by
differentiating Eq.~\eqref{eq:1} with respect to $\kappa$, leading to
\begin{equation}
  \label{eq:2}
  \Kt{\zb}^2 = w \Li{c-1}{\zb\Kt{\zb}} - w \Li{c}{\zb\Kt{\zb}}
\end{equation}
where $d \Li{c}{x} / dx = \Li{c-1}{x}/x$ is used.
For $T>\Tcr$ the pole $\zp$ of $\Zggt{}{}(z)$ is the dominant singularity. It follows from
the equation
\begin{equation}
  \label{eq:3}
  \zp\Kt{\zp} =1\psp.
\end{equation} 
For the calculation of critical exponents,  we always consider the force-free case
characterized by $s=1$  where the RNA is not stretched externally.

\subsection{Scaling of the partition function at $T=\Tcr$}
\label{sec:scal-part-funct}
The order of the branch point exactly at $T=\Tcr$ is calculated by expanding Eq.~\eqref{eq:1} in powers of $z/\zcr -1$ and $\Kt{z}/\Kcr -1$ while keeping $w=\wcr$ fixed. Note that both Eq.~\eqref{eq:2} and~\eqref{eq:3} hold at $T=\Tcr$. One obtains
\begin{equation}
  \label{eq:9}
  \frac{\Kt{z} - \Kcr}{\Kcr} \sim
  - \left(-\frac{z-\zcr}{\zcr}\frac{\zeta_{c-1}}{\Gamma({1-c})}\right)^{\frac{1}{c-1}}\psp.
\end{equation}
Thus, the asymptotic behavior of the generating function at $T=\Tcr$ is
\begin{equation}
  \label{eq:10}
  \Zggt{z}{s=1} = \frac{\Kt{z}}{1-z\Kt{z}} \sim \Kcr\left(-\frac{z-\zcr}{\zcr} \frac{\zeta_{c-1}}{\Gamma(1-c)}\right)^{-\frac{1}{c-1}}\psp.
\end{equation}
$\Zggt{z}{s} \sim C_1(z -\zcr)^{\omega}$, for $z\rightarrow \zcr$ and non-integer
$\omega$, implies $Z_N \sim C_1^{-1} {N^{-(\omega +1)} \zcr}^{-(N+1)}$ and one obtains
\begin{equation}
  \label{eq:11}
    Z_N \sim N^{(2-c)/({c-1})} \zcr^{-(N+1)} \psp.
\end{equation}

\subsection{Expansion of the branch point for $T<\Tcr$}
\label{sec:expans-branch-point}
To obtain the critical behavior for $T<\Tcr$ we perform an asymptotic expansion
of Eqs.~\eqref{eq:1} and~\eqref{eq:2} around the critical point, where the
branch point and the pole coincide. Thus, at the critical point all Eqs.~(1-3)
have to hold and we obtain the critical values exactly as
\begin{equation}
  \label{eq:4}
  \Kcr=\frac12\left(1+\sqrt{1+4w\zeta_c}\right)\psp,\quad\quad \zcr=2\left(1+\sqrt{1+4w\zeta_c}\right)^{-1}\psp,\text{ and}\quad\quad \wcr=\frac{\zeta_{c-1} - \zeta_c}{(\zeta_{c-1} - 2\zeta_c)^2}\psp.
\end{equation}

As an ansatz for $\zb(T)$ and $\Ktb(T)$ we use a power series in
$d=w/\wcr-1$, where $w=\exp[{-\varepsilon/(\kB T)}]$ is the Boltzmann weight of
a base pair. To simplify notation, we define $\alpha = (c-2)^{-1}$ and write
\begin{subequations}
    \label{eq:5}
  \begin{gather}
    \zb/\zcr \sim 1+a_1 d + a_2 d^2 + \ldots + d^\alpha \bigl( a_{\alpha} +
    a_{\alpha+1}d + a_{\alpha+2} d^2+ \ldots \bigr)
    + a_{2\alpha -1} d^{2\alpha -1}+\ldots\\
    \Ktb/\Kcr \sim 1+b_1 d + b_2 d^2 + \ldots + d^\alpha \bigl( b_{\alpha} +
    b_{\alpha+1}d + b_{\alpha+2} d^2+ \ldots \bigr) + b_{2\alpha -1} d^{2\alpha
      -1}+ \ldots
  \end{gather}
\end{subequations}
Plugging the ansatz~\eqref{eq:5} into Eqs.~\eqref{eq:1}
and~\eqref{eq:2} we can solve order by order.  To do so, the series
representation of the polylogarithm $\Li{\nu}{x}$ around $x=1$ is
used~\cite{Erdely1953}
\begin{equation}
  \label{eq:6}
  \Li{\nu}{x} \sim \zeta_\nu - \zeta_{\nu-1} (1-x) + \frac12 (\zeta_{\nu-2} - \zeta_{\nu-1}) (x-1)^2 + \ldots
  (1-x)^{\nu-1} \Bigl(\Gamma(1-\nu) + \frac12 (1-\nu) \Gamma(1-\nu) (1-x) + \ldots\Bigr)\psp.
\end{equation}
We obtain the coefficients for $\zb$
\begin{subequations}
  \begin{align}
    a_{1}&=-\frac{\zeta_{c}}{\zeta_{c-1}}\\
    a_{2}&=\frac{\zeta_{c}^2}{\zeta_{c-1}^{3}}(2\zeta_{c-1} -\zeta_{c})\\
    a_{3}&=-\frac{\zeta_{c}^3}{\zeta_{c-1}^5}
    (5\zeta_{c-1}^2-6\zeta_{c-1}\zeta_{c}+2\zeta_{c}^2)\\
    a_{4}&=\frac{\zeta_{c}^4}{\zeta_{c-1}^7}
    (14\zeta_{c-1}^3-28\zeta_{c-1}^2\zeta_{c}
    +20\zeta_{c-1}\zeta_{c}^2-5\zeta_{c}^3)\\
    a_{5}&=-\frac{\zeta_{c}^5}{\zeta_{c-1}^9}
    (42\zeta_{c-1}^4-120\zeta_{c-1}^3\zeta_{c}+135\zeta_{c-1}^2\zeta_{c}^2
    -70\zeta_{c-1}\zeta_{c}^3+14\zeta_{c}^4)\\
    &\vdots\notag\\
    a_{\alpha}&=0\\
    a_{\alpha+1}&=-\frac{\Gamma({1-c})+\Gamma({2-c})}{\zeta_{c-1}}\left(
      -\frac{\zeta_{c-1}^2-3\zeta_{c-1}\zeta_{c}+2\zeta_{c}^2}
      {\Gamma(2-c)\zeta_{c-1}}
    \right)^{\alpha+1}\\
    a_{\alpha+2}&=\frac{ \bigl(\Gamma(1-c)+\Gamma(2-c)\bigr) \bigl(\zeta_{c-1}^2
      + 4(c-2)\zeta_{c-1}\zeta_{c} + (5-3c)\zeta_{c}^2\bigr)
    }{(c-2)\zeta_{c-1}^3} \left(
      -\frac{\zeta_{c-1}^2-3\zeta_{c-1}\zeta_{c}+2\zeta_{c}^2}
      {\Gamma(2-c)\zeta_{c-1}}
    \right)^{\alpha+1}\\
    &\vdots\notag\\
    a_{2\alpha-1} &= 0\\
    &\vdots\notag
  \end{align}
\end{subequations}
We obtain the coefficients for $\Ktb$
\begin{subequations}
  \begin{align}
    b_{1}&=\frac{\zeta_{c}}{\zeta_{c-1}}\\
    b_{2}&=-\frac{\zeta_{c}^2}{\zeta_{c-1}^{3}}(\zeta_{c-1} -\zeta_{c})\\
    b_{3}&=2\frac{\zeta_{c}^3}{\zeta_{c-1}^5}(\zeta_{c-1} - \zeta_{c})^2\\
    b_{4}&=-5\frac{\zeta_{c}^4}{\zeta_{c-1}^7}(\zeta_{c-1}-\zeta_{c})^3\\
    b_{5}&=14\frac{\zeta_{c}^5}{\zeta_{c-1}^9}(\zeta_{c-1}-\zeta_{c})^4\\
    &\vdots\notag\\
    b_{\alpha}&=-\left(
      -\frac{\zeta_{c-1}^2-3\zeta_{c-1}\zeta_{c}+2\zeta_{c}^2}
      {\Gamma(2-c)\zeta_{c-1}}\right)^\alpha\\
    b_{\alpha+1}&=\frac{\Gamma({1-c})+\Gamma({2-c})}{\zeta_{c-1}} \left(
      -\frac{\zeta_{c-1}^2-3\zeta_{c-1}\zeta_{c}+2\zeta_{c}^2}
      {\Gamma(2-c)\zeta_{c-1}} \right)^{\alpha+1} \notag\\
&\hspace{4ex}    
+\left(
      -\frac{\zeta_{c-1}^2-3\zeta_{c-1}\zeta_{c}+2\zeta_{c}^2}
      {\Gamma(2-c)\zeta_{c-1}} \right)^{\alpha}
    \frac{\zeta_{c-1}^2 - (c-2)\zeta_{c-1}\zeta_{c}-\zeta_{c}^2 }
    {(c-2)\zeta_{c-1}^2}\\
    b_{\alpha+2}&= \frac{\zeta_{c-1} -
      \zeta{c}}{2(c-2)^2\cdot\Gamma(2-c)\zeta_{c-1}^4} \left(
      -\frac{\zeta_{c-1}^2-3\zeta_{c-1}\zeta_{c}+2\zeta_{c}^2}
      {\Gamma(2-c)\zeta_{c-1}} \right)^{\alpha}\notag\\
    &\hspace{4ex}\times
    \biggl(
    2(c-2)\Gamma(1-c)(\zeta_{c-1}- 2\zeta_{c})\bigl( \zeta_{c-1}^2 +
    2(c-2)\zeta_{c-1}\zeta_{c} + (5-3c)\zeta_{c}^2 \bigr)\notag\\
    &\hspace{4ex}\hspace{4ex}
    + \Gamma(2-c)\bigl(
    (c-3)\zeta_{c-1}^3 + (c-3)(4c-7)\zeta_{c-1}^2\zeta_{c} -
    (3c-5)(4c-9)\zeta_{c-1}\zeta_{c}^2 +(3c-5) (4c-7)\zeta_{c}^3 \bigr)
    \biggr)\\
    &\vdots\notag\\
    b_{2\alpha-1}&=-\frac{\zeta_{c-2} -3\zeta_{c-1} + 2\zeta_{c}}{(c-2)\Gamma(2-c)}
    \left( -\frac{\zeta_{c-1}^2-3\zeta_{c-1}\zeta_{c}+2\zeta_{c}^2}
      {\Gamma(2-c)\zeta_{c-1}}
    \right)^{2\alpha-1}\\
    &\vdots\notag
  \end{align}
\end{subequations}

Remarkably, the expansion of the product $\zb \Ktb$ yields
\begin{equation}
  \label{eq:7}
  \zb \Ktb \sim 1+ (a_\alpha + b_\alpha) d^\alpha 
   +(a_{\alpha+1} + b_1a_{\alpha} + b_{\alpha+1} + a_1b_{\alpha}) d^{\alpha+1}+ \mathcal O (d^{\alpha+2}),
\end{equation}
meaning that all integer powers $d^n$, for $n<\alpha$, vanish. 
This fact renders the transition of high order  
when $c \rightarrow 2$ and thus $\alpha \rightarrow \infty$.
For the fraction of bound bases this can be seen directly 
by expanding the low temperature expression
\begin{equation}
  \label{eq:8}
  \theta=2\frac{\Li{c}{\zb\Ktb}}{\Li{c-1}{\zb\Ktb}}
\end{equation}
using Eqs.~\eqref{eq:6} and~\eqref{eq:7}.

\clearpage
\section{Energy parameters}
 The enthalpy $h$ and entropy $s$ parameters used in our calculations are taken
from reference~\cite{Mathews1999}.  For instance, the entries in the row {UA} and
the column {GC},  $ h_{\mathrm{UA,GC}} $ and
$s_{\mathrm{UA,GC}}$,  give the enthalpy and entropy contribution due to the stacking of
the two neighboring base pairs UA and GC, where U and C are located at the
5'-end. The bottom two rows contain the initiation and termination
contribution. For instance, the total free enthalpy of the triple helix
${5'-\mathrm{CGA}-3'}\atop{3'-\mathrm{GCU}-5'}$ is $g_3 = ( h_{\mathrm{CG,GC}}+
h_{\mathrm{GC,AU}} + h^{\mathrm{i}}_{\mathrm{CG}} +
h^{\mathrm{t}}_{\mathrm{AU}}) - T ( s_{\mathrm{CG,GC}}+ s_{\mathrm{GC,AU}} +
s^{\mathrm{i}}_{\mathrm{CG}} + s^{\mathrm{t}}_{\mathrm{AU}}) $.

\begin{center}
  \footnotesize
  \begin{ruledtabular}
    \begin{tabular}{@{}rm{.03em}*{6}{D{.}{.}{2}}m{2em}rm{.0em}*{6}{D{.}{.}{1}}@{}}
      &&\multicolumn{6}{c}{ $\displaystyle{\text{Enthalpy}\   h}\ / \ ({\kilo\calory\per\mole})$}&&&&\multicolumn{6}{c}{ $\displaystyle{\text{Entropy}\   s}\ / \ ({10\rpcubed\kilo\calory\per(\mole\usk\kelvin)})$}\\\cline{3-8}\cline{12-17}
      &&\multicolumn{1}{c}{\text{AU}}&\multicolumn{1}{c}{\text{UA}}&\multicolumn{1}{c}{\text{CG}}&\multicolumn{1}{c}{\text{GC}}&\multicolumn{1}{c}{\text{GU}}&\multicolumn{1}{c}{\text{UG}}&&&&\multicolumn{1}{c}{\text{AU}}&\multicolumn{1}{c}{\text{UA}}&\multicolumn{1}{c}{\text{CG}}&\multicolumn{1}{c}{\text{GC}}&\multicolumn{1}{c}{\text{GU}}&\multicolumn{1}{c}{\text{UG}}\\\hline
      AU&&-6.82&-9.38&-11.40&-10.48&-3.21&-8.81&&AU&&-19.0&-26.7&-29.5&-27.1&-8.6&-24.0\\
      UA&&-7.69&-6.82&-12.44&-10.44&-6.99&-12.83&&UA&&-20.5&-19.0&-32.5&-26.9&-19.3&-37.3\\
      CG&&-10.44&-10.48&-13.39&-10.64&-5.61&-12.11&&CG&&-26.9&-27.1&-32.7&-26.7&-13.5&-32.2\\
      GC&&-12.40&-11.40&-14.88&-13.39&-8.33&-12.59&&GC&&-32.5&-29.5&-36.9&-32.7&-21.9&-32.5\\
      GU&&-12.83&-8.81&-12.59&-12.11&-13.47&-14.59&&GU&&-37.3&-24.0&-32.5&-32.2&-44.9&-51.2\\
      UG&&-6.99&-3.21&-8.33&-5.61&-9.26&-13.47&&UG&&-19.3&-8.6&-21.9&-13.5&-30.8&-44.9\\[1ex]
      $  h^{\mathrm{i}}$&&7.33&7.33&3.61&3.61&7.33&7.33&&$  s^{\mathrm{i}}$&&9.0&9.0&-1.5&-1.5&9.0&9.0\\
      $  h^{\mathrm{t}}$&&3.72&3.72&0.00&0.00&3.72&3.72&&$  s^{\mathrm{t}}$&&10.5&10.5&0.0&0.0&10.5&10.5
    \end{tabular}
  \end{ruledtabular}
\end{center}